%% file: author.tex
\documentclass[graybox]{svmult}

\usepackage{mathptmx}       
\usepackage{helvet}         
\usepackage{courier}        
\usepackage{type1cm}        
\usepackage{makeidx}         
\usepackage{graphicx}        
\usepackage{multicol}        
\usepackage[bottom]{footmisc}

\makeindex             

\begin{document}

\title*{Radiative Transfer in Star Formation: Testing FLD and Hybrid Methods}
\author{James E Owen, Barbara Ercolano and Cathie J. Clarke}
\institute{James Owen \at Canadian Institute for Theoretical Astrophysics, 60 St. George Street, Toronto, Canada \\
Barbara Ercolano \at Excellence Cluster `Universe', Boltzmannstr. 2, 85748 Garching, Germany\\
Cathie Clarke \at Institute of Astronomy, Madingely Road, Cambridge CB3 0HA, U.K.}
%
%
\maketitle

\abstract*{We perform a comparison between two radiative transfer algorithms commonly employed in hydrodynamical calculations of star formation: grey flux limited diffusion and the hybrid scheme, in addition we compare these algorithms to results from the Monte-Carlo radiative transfer code {\sc mocassin}. In disc like density structures a hybrid scheme performs significantly better than the FLD method in the optically thin regions, with comparable results in optically thick regions. In the case of a forming high mass star we find the FLD method significantly underestimates the radiation pressure by a factor of $\sim 100$. }

\abstract{We perform a comparison between two radiative transfer algorithms commonly employed in hydrodynamical calculations of star formation: grey flux limited diffusion and the hybrid scheme, in addition we compare these algorithms to results from the Monte-Carlo radiative transfer code {\sc mocassin}. In disc like density structures the hybrid scheme performs significantly better than the FLD method in the optically thin regions, with comparable results in optically thick regions. In the case of a forming high mass star we find the FLD method significantly underestimates the radiation pressure by a factor of $\sim 100$. }

\section{Introduction}
\label{sec:1}
Numerical models of the star formation process have improved remarkably over the last two decades; however, many questions still remain. In particular the thermal and mechanical feedback produced by the radiation from the forming stars remains to be understood both in low mass, and in particularly  high mass star formation. Several numerical schemes for including the effects of radiation in hydrodynamic codes have been proposed including: Monte-Carlo techniques (Harries 2011); Short Characteristics (Davies et al. 2012); Flux limited diffusion and other moment methods (Levermore \& Pomraning 1981); Pure ray-tracing techniques (Abel \& Wandelt 2002) and Hybrid techniques - which combine various method together to arrive at a hopefully improved and faster method - (Edgar \& Clarke 2003; Kuiper et al. 2010). While all these methods are fast enough for inclusion into a hydrodynamical calculation, the algorithm which is both fast and accurate for incorporation into a star formation calculation still remains a matter for debate. In this work I will present radiative transfer tests of the most commonly used method: Flux Limited Diffusion (FLD), and its improvement in the form of a hybrid method.  

\section{The Flux Limited Diffusion and Hybrid Schemes}
\label{sec:2}
The flux limited diffusion approximation attempts to simplify the radiative transfer problem by using a moment method, where the radiative transfer variables are replaced by angle averaged quantities, which are the only ones required in a hydrodynamical calculation. To do this a flux limiter ($\lambda$) is employed which allows the radiative transfer problem to be written as a diffusion equation:
\begin{equation}
\frac{\partial E_\nu}{\partial t}+\nabla\cdot\left(\frac{c\lambda_\nu}{\kappa_\nu\rho}\nabla E_\nu\right)=S_\nu
\end{equation} 
where $E$ is the internal energy and $S$ is any appropriate source terms. In this form $\lambda$ has the limits $\rightarrow 1/3$ in optically thick media, and $\rightarrow \kappa\rho E/\nabla E$ in optically thin media. Thus the flux limiter ensures the radiation field is transported at the correct speed in both optically thick/thin limits. A common choice is to also employ the FLD method in a grey approximation, meaning the radiation field is in local thermal equilibrium with the matter. 

In reality the radiation field is not always locally thermalised, and the temperature and radiation pressure are correctly given by:
\begin{eqnarray}
T=\left(\frac{\kappa(T_{\rm source})}{\kappa(T)}\frac{E}{a}\right)^{1/4} &, &\,\, a_{\rm rad}=\frac{\kappa(T_{\rm source})}{c}F 
\end{eqnarray}  
however, in the grey case they are replace by:
\begin{eqnarray}
T=\left(\frac{E}{a}\right)^{1/4} &, &\,\, a_{\rm rad}=\frac{\kappa(T)}{c}F 
\end{eqnarray}  
where $T_{\rm source}$ is the radiation temperature the opacity source sees, $a$ is the radiation constant and $F$ the flux. Thus, in an optically thin region $T_{\rm source}$ does not in general equal $T$ and in the case the source is the photosphere of a YSO then it is typically much greater than the local temperature. Meaning the temperature and radiation pressure will be underestimated by an amount depending on $\kappa(T_{\rm source})/\kappa(T)$, which can be very large in the case of a massive star. Furthermore, this has an important feedback on the flux limiter which depends on $\kappa(T)$ in the grey approximation, thus if T is underestimated then so is $\kappa$ (for standard dust opacities) leading the flux-limiter to become optically thin too easily. 

A simple improvement upon this is a hybrid approach to deal with the source radiation field (e.g. Edgar \& Clarke 2003, Kuiper et al. 2010), which uses a multi-frequency ray-tracing approach to deal with the direct radiation from a forming YSO and then a grey solver (e.g Diffusion approximation - Edgar \& Clarke or FLD - Kuiper et al. 2010) to deal with the re-radiated fields. In the hybrid  method the directly attenuated flux just becomes part of the source terms in Equation~1. This means that one is now using a frequency dependant approach when the photosphere of the YSO can be seen and hence when errors resulting from large $\kappa(T_{\rm source})/\kappa(T)$ would be greatest are nullified. 

\section{Comparison and tests}

In order to asses the relative accuracy  of the   classical FLD method versus the hybrid method we perform two benchmark calculations one where thermal pressure dominates, as an example of a disc around a low mass star; secondly we consider a calculation where the radiation pressure force dominates, as an example of a disc around a forming high mass star. We compare the results of the two benchmark calculations performed with the FLD method, the hybrid method and the full Monte Carlo radiative transfer using the {\sc mocassin} code (Ercolano et al. 2003, 2005).  We adopt a density profile of the form:
\begin{equation}
\rho=\rho_0\left(\frac{R}{R_0}\right)^{-2}\exp\left(-\frac{z^2}{2H(R)}\right)
\end{equation}
where $H(R)$ is the scale height which is taken to be  flaring of the form $H(R)=(H/R)_0R^{1.1}$. 

For the low luminosity star we adopt a disc mass of $0.05$M$_\odot$, a stellar luminosity of 1L$_\odot$ with a photospheric temperature of $4000$K. In the case of the high luminosity star we adopt a disc mass of $1$M$_\odot$, a stellar  luminosity of $10^6$L$_\odot$ and a photospheric temperature of $20,000$K.  We find that in the case of the low mass star the temperature differences between the FLD, Hybrid and {\sc mocassin} results are small, in the optically thick mid-plane of the disc, where the differences are at the few percent level. In the optically thin region the Hybrid and {\sc mocassin} results agree perfectly and the FLD method underestimated the temperature by a factor of $\sim 2$, arising from the errors in $\kappa(T_{\rm source})/\kappa(T)$ discussed above. 

In the high mass, radiation pressure dominated case the differences are more severe, in the optically thick mid-plane the FLD method gives the lowest temperature with a $\sim 20-30\%$ difference to the {\sc mocassin} results and the Hybrid method gives a slightly lower temperature than the {\sc mocassin} results with a ~5-10\% difference. In the optically thin cavity the {\sc mocassin} and Hybrid results agree perfectly,  whereas the FLD method underestimates the temperature by a factor $4-5$, misplaces the radius of the dust destruction front  (at $T=2000$K) by a factor of $\sim 3$ and severely underestimated the radiation pressure - which is the dominant force term - by a factor of $\sim 100$, as shown in Figure~1.    

\section{Conclusions}
We have performed comparisons between the FLD and Hybrid radiative transfer schemes in the cases of discs around low-mass and high mass stars. In both cases we find that the hybrid scheme performs better than the FLD scheme when compared to Monte-Carlo calculations, although in the case of low mass stars where radiation pressure is unimportant the FLD adequately describes the disc temperature where most of the dynamically important material is. However, in the case of discs around high mass-stars we find FLD to lead to large errors in both the temperature and radiation. . We conclude that the hybrid scheme is more suited
to simulations involving massive stars and that FLD
should be used with caution in cases where the optically thin region or interfaces between optically thin and
optically thick regions are dynamically important.

%
\begin{figure}
\centering
\includegraphics[width=0.85\textwidth]{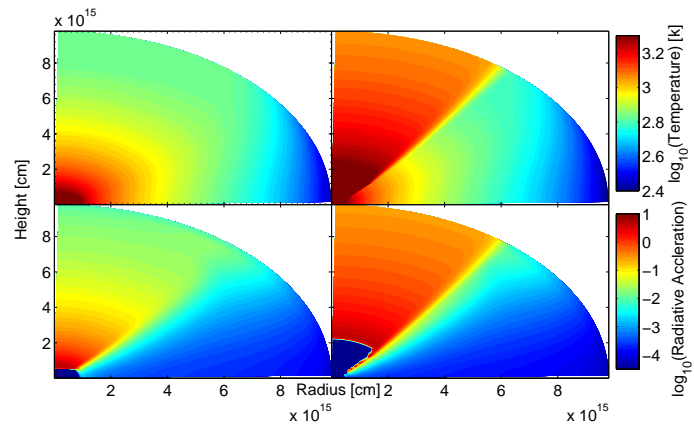}
\caption{Comparisons between the FLD method (left panels) and the Hybrid method (right panels) for the temperature structure (top panels) and radiation pressure (bottom panels) of a disc around a forming high mass star. The dark blue regions in the bottom panels show the dust destruction front.}
\end{figure}
\input{referenc}

\end{document}

%% file: referenc.tex
%
%
%